\documentclass[twocolumn,prd,floatfix]{revtex4}
\usepackage{graphicx}
\begin{document}


\title{Cosmological N-Body simulation: Techniques, Scope and Status} 


\author{J.~S.~Bagla}


\email{Email: jasjeet@mri.ernet.in}  


\affiliation{Harish-Chandra Research Institute, Chhatnag Road,\\
Jhunsi, Allahabad 211 019, India.}


\begin{abstract}
Cosmological N-Body simulations have become an essential tool for studying 
formation of large scale structure.  
These simulations are computationally challenging even though the available
computing power gets better every year.  
A number of efficient algorithms have been developed to run large
simulations with better dynamic range and resolution. 
We discuss key algorithms in this review, focusing on techniques used and
their efficacy. 
N-Body simulations solve a model that is an approximation of the
physical model to be simulated, we discuss limitations arising from
this approximation and techniques employed for solving equations.
Apart from simulating models of structure formation, N-Body
simulations have also been used to study aspects of gravitational
clustering. 
Simulating formation of galaxies requires us to take many physical
process into account; we review numerical implementations of key
processes.   
\end{abstract}


\maketitle


\section{Introduction}

Large scale structures like galaxies and clusters of galaxies are believed to
have formed by amplification of small perturbations
\cite{lssu,tp93,peebles,peacock,colesbook,tpvol3,dodelson,liddlebook,
  lss_review}.  
Galaxies are highly over-dense systems, matter density $\rho$ in galaxies is
thousands of times larger than the average density $\bar\rho$ in the
universe.  
Typical density contrast ($\delta \equiv \rho/{\bar\rho} - 1$) in matter at
these scales in the early universe, e.g. at the time of decoupling of
matter and radiation was much smaller than unity.  
Thus the problem of galaxy formation and the large scale distribution of
galaxies is essentially one of evolving density perturbations from small
initial values to the large values we encounter today.

The universe is assumed to be homogeneous and isotropic at large scales and
this assumption is consistent with observations and there are strong
limits on departures from homogeneity and isotropy \cite{glob_prop}. 
A homogeneous and isotropic universe is described by Friedman equations in
the general theory of relativity.   
As long as perturbations in the gravitational potential are small, we can
treat density fluctuations as perturbations about a Friedman universe. 
If perturbations are in non-relativistic matter, as appears to be the case in
our universe, we can work in the Newtonian limit for studying their
evolution. 
The back reaction of perturbations on the universe is not taken into account,
i.e., we do not worry about the effect of perturbations on the average global
properties of the universe. 
Studies have shown that this back-reaction is ignorable in most cases
\cite{avrg1,avrg2}. 
Local variations in density, etc. can lead to dispersion in values of
cosmological parameters determined through local observations
\cite{hubble_dist} but this problem can be controlled by making measurements
at larger scales. 

Gravity is the dominant force at large scales and is believed to drive growth
of perturbations. 
Magnetic field is the only other interaction that can lead to
formation of large scale structures, this leads to very distinct
signatures and does not appear to be the dominant factor in our
universe \cite{mag1,mag2,mag3}. 
We will assume that gravity is the only interaction responsible for
growth of perturbations at large scales. 
Equations that describe the evolution of density perturbations in
non-relativistic matter due to gravitational interaction in an expanding
universe have been known for a long time \cite{deleqn}. 
These equations can be solved analytically for small density contrasts, and
for highly symmetric situations.
But apart from such special cases, few solutions are known.   
Many approximate solutions are known
\cite{za,adh,adh2,ffa,lep,fpa,approxrev1,approxrev2,lss_review} and are useful
in understanding the evolution of perturbations in the
quasi-linear regime, these fail when density contrast becomes large
($\delta \gg 1$).  
Cosmological N-Body simulations are an essential tool for evolving density
perturbations in the non-linear regime. 
Fluctuations in the gravitational potential do not grow by a large amount even
as density contrast increases by several orders of magnitude \cite{lep,fpa},
therefore the Newtonian approximation continues to be a valid framework. 
At galactic scales, gas dynamical and other effects play an important role and
need to be taken into account for a detailed solution of the problem.

N-Body simulations are used for a variety of applications.  
Simulations of specific models of dark matter allow us to make predictions for
these models and compare with observations.
Simulations allow us to carry out numerical experiments with initial
conditions that have little to do with the real universe.  
The purpose of such experiments is to understand the physics of gravitational
collapse in an expanding universe. 
Simulations are also used for testing approximate solutions for growth
of density perturbations, comparisons with N-Body simulations allow us
to validate these approximations and understand when these
approximations are useful. 
Lastly, we can calibrate methods for analysing observations on mock catalogues 
made from N-Body simulations. 
We can test whether a particular method works or not because in N-Body
all the details are known whereas the same is not true of the real universe. 

The physical parameters of the problem make cosmological simulations a 
challenging task.  
Unlike simulations of systems like globular clusters that can be treated as
(relatively) isolated systems, the universe does not have a boundary.  
Therefore cosmological simulations need to be run with periodic boundary
conditions.  
An exception are simulations of large spherical volumes that do not suffer
significant deformation during the course of evolution.

\begin{figure}
\includegraphics[width=3.4truein]{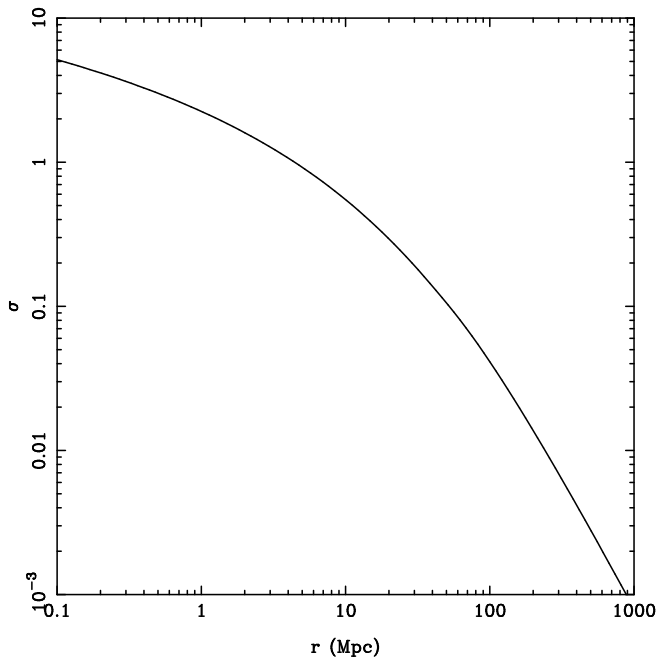}
\caption{Root mean square fluctuations in mass $\sigma$ are shown as a
  function of scale.  The amplitude of $\sigma$ is for the $\Lambda$CDM model
  with $\Omega_\Lambda=0.7$, $\Omega_B=0.05$, $h=0.7$ and $n=1.0$.  The
  linearly extrapolated amplitude is plotted here and no non-linear
  corrections have been included in generating this plot.}
\label{sigma}
\end{figure}

Observations suggest that density perturbations are present at
all scales that have been probed by observations.  
Amplitude of fluctuations at small scales is large and it drops at large
scales ($l \gg 10$~Mpc where $1$~Mpc = $3.08 \times 10^{24}$~cm.).
Figure~\ref{sigma} shows the root mean square fluctuations in mass as a
function of length scale for one of the popular models. 
Fluctuations at scales much larger than $100$~Mpc are generally not relevant
for structure formation at the scale of galaxies.   
Thus the physical size corresponding to the periodic simulation box 
should be at least $100$~Mpc, unless these are meant for studying
large scale structure at early times when the amplitude of fluctuations is
small and a smaller simulation box is acceptable. 
We will revisit this issue and discuss it more quantitatively but this
approximate figure will suffice at present.

If we wish to study the distribution of galaxies in detail then the mass of
each particle in the simulation should be much smaller than the mass of a
typical galaxy.  
This, with the requirement that the simulation box should be more than
$100$~Mpc across implies that the N-Body simulation must be done with
at least $10^8$ particles.   
The large number of particles required is one of the things that makes
cosmological simulations challenging.  

Observations show that the dominant component of matter in the universe does
not radiate light and thus cannot be seen, except through its gravitational
effect on visible matter \cite{dm_review}. 
Observational evidence points towards non-relativistic dark matter
\cite{wdm,wmap,wmap_sdss}.  
This makes our task simple as motions of non-relativistic matter can be
studied in the fluid limit at large scales. 
(Relativistic dark matter will have significant pressure/velocity dispersion
and one cannot take the fluid limit.  
In such a case one needs to solve the Boltzmann equation in order to correctly
model the effects of free streaming\cite{neutrino_ma}.) 
Little is known about what constitutes this dark matter though it is often
assumed to be made of {\sl Weakly Interacting Massive Particles (WIMP)}.
There are strong limits on the interaction cross section of WIMPs from
astrophysical considerations \cite{idm}.  
Non-relativistic, non-interacting (collisionless) dark matter is known as {\sl
  Cold Dark Matter}\/ (CDM) \cite{cdm1,cdm2,cdm_review}.  
Dark matter interacts only through gravity and that makes simulating growth of
perturbations somewhat simpler. 

Dark matter dominates over normal matter in terms of density by a large
factor \cite{wmap}, therefore N-Body simulations with the entire matter
density in dark matter provide a good first approximation for the
distribution of matter.   
In any case, gravity is the dominant interaction and normal matter is
expected to follow dark matter at large scales. 

Cosmological simulations differ from other types of N-Body simulations
as the background is not static and expansion of the universe has
to be taken into account.
It is convenient to work in {\sl comoving} coordinates that expand with the
universe.
Equations in comoving coordinates deal with the evolution of
perturbations and the average quantities (density, velocity, etc.) are
scaled out.   
We will discuss this in greater detail in the following section.

In N-Body simulations, each particle represents a very large number of
dark matter particles and interaction of two particles in N-Body
simulations should mimic the interaction of two ``fluid elements''.
The fluid elements being simulated have a physical size and at scales
comparable to this, the fluid elements should feel much less force
than two point particles. 
This is done by assuming that the particles have a finite size and
density profile, this leads to an effective softening of force at
small scales. 
Clearly, the force should be softened at scales comparable to the
(local) average inter-particle separation in the N-Body simulation.  
A much smaller softening length can lead to unwanted two body
relaxation \cite{twobody,obliquewave}.
The form of force softening also plays an important role, force
softened in such a way that it matches inverse square force beyond the
softening length is better \cite{optsoft1,optsoft2}.  
If the softened force approaches inverse square force only
asymptotically then the difference is equivalent to error in force at
large separations and can lead to spurious effects.  

This ends the overview of the physical requirements and associated
approximations for cosmological simulations.
The next section contains a detailed discussion of cosmological N-Body
simulations that take only gravitational interaction into account.  
It is very important to understand limitations of N-Body simulations as using
simulations outside the domain of validity can easily lead to incorrect 
results.  
Limitations of N-Body simulations are discussed in the following section.  
N-Body simulations have been used to further our understanding of
gravitational clustering in the non-linear regime, we discuss some
relevant issues. 
This is followed by an overview of other processes that must be taken into
account for a complete study of galaxy formation, we also discuss
numerical implementations of these physical processes in N-Body simulations.


\section{Gravity Only Simulations}

In this section we discuss N-Body simulations where only gravitational
interaction is taken into account.  
In an N-Body simulation a given density-velocity field is represented
by a set of particles \cite{sim_book}.   
Density as a function of position is obtained by averaging over this
distribution of particles.  

Evaluation of force and solving the equation of motion are two key
components of N-Body simulations. 
Setting up relevant initial conditions for cosmological simulations is
another important aspect. 
We discuss the equation of motion and its integration in the following
subsection.  
This is followed by a discussion of algorithms for force calculations.
We end this section with a discussion on setting up the initial
conditions. 


\subsection{Equation of Motion}  

The equations that govern the evolution of a given distribution of particles
are obtained by starting with the standard (Newtonian) equation of motion for
a set of gravitationally interacting particles in the physical/proper
coordinates. 
Transforming to comoving coordinates allows us to focus on perturbations in
density and velocity.
Information about expansion of the universe appears in these equations
through the scale factor $a(t)$, obtained as a solution to the Friedman
equations
\cite{lssu,tp93,peebles,peacock,colesbook,tpvol3,dodelson,liddlebook}.  
The scale factor and other cosmological parameters in these equations
carry information about the average quantities that are scaled out in
the coordinate transformation.  
\begin{eqnarray}
\ddot{{\mathbf x}} + 2 \frac{\dot{a}}{a} \dot{{\mathbf x}} &=& - \frac{1}{a^2}
\nabla\phi \label{motion} \\
\nabla^2\phi &=& 4 \pi G {\bar\rho}(t) a^2 \delta = \frac{3}{2}
H_0^2 \Omega_0 \frac{\delta}{a}  \label{poisson}
\end{eqnarray} 
Here ${\mathbf x}$ is the comoving position of a particle and is related to
the physical/proper position ${\mathbf r} = a(t) {\mathbf x}$, with $a(t)$
being the scale factor. 
$\phi$ is the gravitational potential due to density perturbations, $H_0$ is
the Hubble's constant and $\Omega_0$ is the density parameter of
non-relativistic matter at the present epoch
\cite{lssu,tp93,peacock,colesbook,tpvol3,lss_review}.   
It is assumed that the relativistic components do not cluster, or are
negligible in our universe.
These equations are valid for non-relativistic matter ($v \ll c$, $\phi \ll
c^2$) at scales that are much smaller than the Hubble radius ($l \ll c/H_0$)
\cite{lssu,tp93}. 

The expansion of the universe acts as a viscous force in comoving
coordinates.  
This drag opposes gravitational infall and as a result the growth of
density perturbations is slower in an expanding universe. 
The time scale over which gravitational infall occurs (in absence of
expansion) is comparable with the expansion time scale ($a / {\dot a}$),
therefore velocities of particles do not become very large during infall. 
As a result integrating equation of motion is simpler in cosmological N-Body
simulations.

The basic idea for numerical integration is as follows. 
The equation of motion expresses the second derivative of position in
terms of position, velocity and time.    
Position and velocity at later times are expressed in terms of
position and velocity at earlier times using a truncated Taylor
series.  
The simplest truncation is not sufficiently accurate and the resulting
error is of order $h^2$ in one time step, where $h$ is the time step
\cite{sim_book,antia,numrec}.   
The key constraint in cosmological simulations is that force
evaluation is very time consuming and one wishes to minimise the
number of force evaluations per time step.  
Mainly due to this reason, cosmological N-Body simulations use the
Leap-Frog method \cite{sim_book,antia,numrec} for integrating the
equation of motion as it requires only one evaluation of force and the
error is of order $h^3$.  
Performance of the Leap-Frog integrator can be improved considerably
by making small modifications \cite{betleap}, but such
modifications are often more useful in non-cosmological N-Body
simulations.  

Time step $h$ is typically chosen so that momentum is conserved and
energy evolves according to the Irvine-Layzer equation
\cite{irvine,layzer,zel_energy}.  
Monitoring consistency with the Irvine-Layzer equation requires care
and adds significantly to the number of operations to be carried out
in an N-Body simulation \cite{p3m}, hence it is usual to carry out
test runs and fix the value of $h$.  
Additional tests can be devised, e.g. we can require that the
clustering of power law models evolves in a scale invariant manner
even in the strongly non-linear regime.

Optimum value of time step $h$ depends on the distribution of
particles and it changes as this distribution evolves. 
It is common to use a time step that varies with time so that the
N-Body code does not use too small a time step when a larger value
will do, or use too large an $h$ when a smaller value is required for
conserving integrals of motion.  
It is possible to generalise even further and choose a different $h$
for each particle as well, motivation for this being that a few
particles in a very dense regions require a small $h$ whereas most
particles are not in such regions.  
There are several methods of implementing this in N-Body simulations,
e.g. see \cite{gadget}. 
Main consideration is to ensure that the positions and velocities of
all particles are synchronised at frequent intervals. 
Using individual time steps can speed up N-Body simulations by
a significant amount.


\subsection{Calculating Force}

Gravitational force in the Newtonian limit falls as $1/r^2$, hence it
is a long range force and we cannot ignore force due to distant
particles.  
This makes calculation of force the most time consuming task in N-Body
simulations.  
As a result, a lot of attention has been focused on this aspect and many
algorithms and optimising schemes have been developed.
We will discuss the major algorithms in some detail and briefly
summarise other developments.   
We refer the reader to \cite{bertsc98} for a detailed review. 
In the following discussion, we also review some algorithms that are
not used in cosmological N-Body simulations as these can be used in
hybrid algorithms.  


\subsubsection{Direct Summation or Particle-Particle Method}

The most obvious approach to the problem of force calculation is to
carry out a direct pairwise summation over all particles.  
This is also called the Particle-Particle (PP) method and this works
very well for a small number of particles. 
Most early simulations used this method, see \cite{nbody_hoyle} for an
early approach, \cite{direct_char} for a discussion of characteristics
of this method, and, \cite{aarseth_review} for an overview of the
direct summation method. 
Earliest cosmological simulations also used this method
\cite{aarseth_cosmo}.  
The number of terms in the pairwise summation increases in proportion
with $N^2$, where $N$ is the number of particles.   
This rapid variation limited the early cosmological simulations to
about $10^3$ particles. 
Limitations of the PP approach were realised as focus shifted to other
methods.  

The last decade has seen a revival of sorts for this method with the
advent of the GRAPE chip \cite{grape}.
With the latest version of GRAPE \cite{grape6} it is possible to
simulate systems with more than $10^6$ particles.  
Development of efficient parallel algorithms \cite{pardirect} has also 
made this method competitive.

It is difficult to implement periodic boundary conditions in the PP method. 
The only optimised method available for this is the Ewald summation
\cite{ewald}.   
One of the first concrete proposals to use this method in cosmological N-Body
simulations is \cite{ewald_r}, though Ewald summation had been used for
testing other methods \cite{p3m}. 
Adding periodic boundary conditions to a PP code remains a difficult
proposition and this method is not used very often for cosmological N-Body
simulations. 


\subsubsection{Tree Method}

The key limitation of the PP method is the rapid increase in
computational load with the number of particles in the N-Body
simulation. 
This in turn arises from adding individual contribution to the force
due to each particle. 
The force of a distant group of particles can be approximated by the
force due to a single pseudo particle located at the centre of mass of
the group, with mass equal to the total mass of the group of
particles. 
This approximation changes the scaling of the number of calculations
from $N^2$ to $N\log N$.  

Efficient division of particles into groups can be done by arranging
particles in a tree structure \cite{bh86,binary_tree}. 
The simulation volume is taken to be a cube and is divided into
smaller cubes with $1/8$ the volume each at every stage till the smallest
cells have only one particle in them. 
Larger cells serve as groups of particles for a rapid calculation of
force. 
An essential ingredient is the criterion for deciding whether a group of
particles can be considered distant or not.
This is called the cell acceptance criterion and the error in
approximation is controlled by the choice of this criterion
\cite{bh86,skel,gadget}
See \cite{skel,perf_tree,perf_tree1,gadget} for a detailed study of
characteristics of the tree code, in particular of errors and timing
as a function of the distribution of particles and the cell acceptance
criteria. 

Accuracy of the tree approximation can be improved by retaining
information about moments of the particle distribution in the
group, e.g. the quadruple moment. 

The tree code can be optimised by vectorising the calculation of force
\cite{vectree_hernquist,vectree_makino}. 
The set of acceptable distant groups of particles is almost the same
for neighbouring particles and sharing of interaction lists amongst
neighbouring particles can further improve the performance of the tree
code \cite{grp}.  
The tree code can be parallelised efficiently \cite{treepar96},
and a parallel tree code has also been implemented using the GRAPE chip
\cite{jm_partree}.  
The parallel algorithm divides the simulation box into domains with equal
number of particles and calculations for each domain are done by a
different processor. 
This scheme can be improved by dividing into domains with equal
computational load \cite{treepar96}. 

Periodic boundary conditions are difficult to implement with tree
codes, the level of difficulty being similar to that with the PP
codes. 
Some innovative schemes have been tried \cite{qperi}.
Implementations using the Ewald summation \cite{ewald} add a large
computational overhead \cite{ewald_h,gadget}.
But in spite of these difficulties, tree codes have been used very
effectively for cosmological N-Body simulations.  


\subsubsection{Fast Multipole Method}

The performance of tree codes can be improved upon by using including
higher moments of mass distribution in cells.
These and use of some other optimisation schemes leads to the fast
multipole method.  
The number of computational operations in the fast multipole method
\cite{multipole} scales as $N$, the number of particles.  

Inclusion of higher moments can be also modelled in terms of {\sl
  pseudo-particles} for easy implementation on the GRAPE chip
  \cite{pseudopart1,pseudopart2}.   

An explicitly momentum conserving extension of the tree code has also
been proposed and implemented \cite{momcons,ordern}, here the number
of computations required scales linearly with the number of
particles. 

These codes also suffer from the problem of open boundary conditions
and cannot be adapted very easily to cosmological problems.   


\subsubsection{Particle-Mesh Method}

Particle-Mesh (PM) method
\cite{negtime,miller,sim_book,p3m,pm_model,bouchet_pm,jbtp_pm} has been used
extensively for cosmological simulations and was the first method to
be used for ``large'' ($N \sim 10^5$) simulations \cite{negtime}.  
In PM codes, the fact that the Poisson's equation (eqn.\ref{poisson})
is a simple algebraic equation in Fourier space is combined with Fast
Fourier Transforms (FFT) \cite{antia,numrec}.  
FFT requires sampling of functions at uniformly spaced points, and a
grid/mesh is used for this.  
Usually the simulation volume is taken to be a cube with equal number of
grid/mesh points along each axis. 

Particles are used for representing the density and velocity field 
and we compute density at grid points by using weight functions
\cite{sim_book}. 
Density contrast $\delta$ and the potential $\phi$ is defined on the
grid for solving the Poisson equation.
Use of particles and a mesh gives this method the very appropriate
name.  

By using Fourier methods we get periodic boundary conditions for free. 
The use of a mesh also softens the force naturally at small scales, though
this softening leads to underestimation of force to fairly large scales
\cite{pm_model,bouchet_pm}.  
Thus it would seem that PM codes are ideal for cosmological N-Body
simulations if we wish to study the large scale structure of the
universe. 

The PM algorithm has been parallelised \cite{pmfast}, typically this
involves using parallel FFT \cite{fftw}. 
The computational load is divided amongst processors by dividing the
simulation box into slabs of equal size.

Softening at the mesh scale ensures collisionless evolution, but this
also means that PM codes cannot resolve structure at length scales
smaller than a mesh. 
Even in dense regions force softening is at the grid scale. 
This seriously limits the effective dynamic range of simulations run
with PM codes. 
A related problem is that the use of mesh makes force of a particle
anisotropic at small scales.  

Improving force resolution in high density regions can improve the
effectiveness of PM codes.   
Several techniques have been proposed for achieving this as poor
resolution is the main shortcoming of PM codes for cosmological N-Body
simulations.  


\subsubsection{Adaptive Mesh Refinement}

In {\sl Adaptive Mesh Refinement} (AMR) the grid is refined in high density
regions. 
A new mesh with smaller spacing is introduced and the low resolution force
calculated using the coarse global grid is improved upon using the refined
mesh \cite{amr_villumsen,amr_suisalu,amr_klyp,mlapm}. 
Several levels of refinement can be introduced; indeed are required in order
to resolve substructure in dense haloes \cite{amr_klyp,mlapm}.  
Care is required to ensure conservation of momentum and angular momentum in
AMR codes. 


\subsubsection{P$^3$M: Particle-Particle + Particle-Mesh}

The basic idea here is to add a ``correction'' to the force
computed using the PM method.  
This correction is computed by summing the contribution of close neighbours
using the particle-particle method, hence the name PP$+$PM $=$ P$^3$M.   
It is assumed that this correction depends only on the distance, i.e., it is
assumed to be isotropic, and is generally added out to a distance of about
$1.5-2$ times the distance between neighbouring mesh points\cite{p3m,cman}.  
P$^3$M was the first method to be used for high resolution cosmological N-Body
simulations.  

The PP part of the calculation can also be done with the GRAPE chip
\cite{grape_p3m}, though it requires some innovation as the chip is designed
to return the force and not the short range correction.  

The P$^3$M has been parallelised, though load balancing for such a code is not
very simple as the overdense regions that require more CPU time are not
distributed uniformly \cite{pp3m,parp3m,p4m} (see below). 

The P$^3$M has some undesirable features.
\begin{itemize}
\item
The correction for the force is assumed to be isotropic, whereas the
standard PM force has anisotropies at grid scale due to the
anisotropic mesh structure.   
Thus the resulting force (long range $+$ short range correction) must be
anisotropic at the grid scale.   
\item
The short range correction in force is added only up to $1.5-2$ grid lengths,
whereas the PM method underestimates the force out to a much larger
distance\cite{bouchet_pm}.  
\item
The refined inter-particle force is softened at scales much smaller than the
average inter-particle separation, this can lead to two body scattering and
relaxation \cite{twobody,obliquewave}.  
Results of astrophysical interest like the correlation function may also get
modified in the process \cite{discr3}.   
\item
P$^3$M simulations slow down at late times when the distribution of particles
becomes highly clustered.
At this stage computation of the short range correction of force dominates the
total number of compute operations required.  
\end{itemize}

While the last item here relates to the efficiency of the code, other items in
the list are far more serious as these raise doubts about the accuracy with
which force is calculated. 
In order to retain the good features of P$^3$M codes and address some of the
issues listed above, several variations on the theme have been suggested.  


\subsubsection{Tree + PM = TPM, GOTPM, TreePM . . .}

In this section we discuss a series of hybrid codes that combine the PM and
the tree method in the same spirit as the PP and PM methods are combined in
the P$^3$M code. 
We will discuss these in order of the significance of departure from the
P$^3$M method. 

\begin{itemize}
\item
The {\sl Grid Of Trees PM} (GOTPM) code \cite{gotpm} replaces the PP part of
P$^3$M codes with a local tree in each region.
This speeds up calculation of the short range force correction and the time
taken for this calculation is less sensitive to the degree of clustering.
This code takes care of the last undesirable feature listed for P$^3$M codes
but it does not address any of the other issues. 
GOTPM is a parallel code where multiple levels of decomposition is used
to achieve load balance \cite{gotpm}.
\item
The TPM code \cite{tpm,tpmn,tpmnew} also addresses the problem of two body
scattering in P$^3$M codes.  
In this code the short range correction to force is added only if the particle
is in a high density region. 
Density is computed at the position of each particle and a local tree is
constructed in high density regions for computing the short range correction
to the long range PM force.  
If the high density regions have a large number of particles then the region
is fragmented into sub-regions.  
TPM is a parallel code, ideally suited for distributed parallel computing. 
\item
The {\sl Tree $+$ PM} (TreePM) code \cite{treepm} makes
significant changes to the P$^3$M approach.  
The force is partitioned between a short range and a long range component
instead of using the force computed with a PM code as the long range force. 
In the TreePM code the long range force is truncated below a certain
scale $r_s$ and the long range force becomes very small below this scale. 
In particular the long range force is small at grid scale if $r_s$ is chosen
appropriately and therefore anisotropies in the long range force are also less
important compared to the P$^3$M, GOTPM and TPM codes. 
The short range force is computed using a global tree. 
The short range force has to be taken into account out to larger distances
than in a typical P$^3$M code. 
By tuning the model parameter $r_s$ and a few other parameters in the code, it
is possible to keep error in force below $1\%$ for most of the particles
\cite{error_treepm}. 
The TreePM code is fairly simple to implement as the mathematical model of
this method is well defined \cite{error_treepm,mwhite_treepm,gadget2}.   
This code has been parallelised in a manner similar to that used for tree
codes with some provision for the long range force calculation
\cite{treepm_par}.   
The TreePM code solves all the other problems of a P$^3$M code but it does not
address the issue of two body scattering and relaxation.
\item
The Adaptive TreePM (ATreePM) code \cite{atreepm} addresses the problem of two
body scattering in TreePM code by using an adaptive softening length instead
of a constant softening length.  
This can be thought of as an equivalent of AMR without using a refined grid.  
Local number density of particles is used to determine the softening length
for particles. 
In order to ensure momentum conservation, force is symmetrised for particles
that are closer than the softening length of either one of the two particles. 
Determination of local density and explicit symmetrisation add a computational
overhead. 
This is offset to some extent by the speedup at early times when the
softening length is large for all the particles.  
Using a hierarchy of time steps optimises this method further.
\end{itemize}

The variety of techniques used for computing gravitational force discussed
above is evidence of the work being done in developing better algorithms.  
It is also evident that full use has been made of parallel computing in order
to achieve good performance. 

It is important to construct ways of comparing performance of
different N-Body codes.  
We list some suggestions here.
\begin{enumerate}
\item
Dynamic range: The range of scales over which interaction force is computed
reliably. 
There is rarely any problem at larger scales so the dynamic range is
typically determined by the smallest separations over which the force
errors are small. 
A useful unit for this scale is the average inter-particle separation.
We can fix this scale by requiring that at all larger scales error in
force due to one particle be less than $1\%$.
For good codes, this scale should coincide with the softening length. 
\item
Trajectories of particles: The code should integrate the equation of
motion in a reproducible manner and momentum should be conserved.  
The N-Body code should reproduce well known results about the correlation
function and other statistical measures in the quasi-linear regime. 
Time step should be much smaller than the crossing time of particles
in dense haloes, the ratio of time step to the smallest crossing time
is a good estimate and this number should be smaller than $10^{-1}$.
\item
Efficiency: The N-Body code should be efficient and we should be able to run
large simulations in as little time as possible. 
This requirement is likely to conflict with the first two
requirements, and one should compare both the dynamic range and
efficiency.   
It has been proposed that error in force should be plotted against
time taken for computing force for comparing codes \cite{gadget}, we
believe this to be the correct approach.
\item
Resource requirement: The N-Body code must store positions and
velocities of all the particles, i.e., at least $6$ numbers per particle. 
Most advanced codes discussed here store many more numbers per particle in
order to speed up force calculation.  
This requirement can restrict our ability to run large simulations as the
memory on computers is limited.  
In case of parallel codes, the relevant quantity is the maximum memory
utilisation per particle for one processor. 
\end{enumerate}

This ends the discussion of different algorithms for computing force in
cosmological N-Body simulations. 


\subsection{Setting up Initial Conditions} 

N-Body simulations are generally started from fairly homogeneous
initial conditions, i.e. the density contrast is much smaller than
unity at all scales of interest in the simulation. 
In this regime we can use linear perturbation theory to compute all
quantities of interest.
In linear theory, the evolution of density contrast can be described
as a combination of a growing and a decaying mode.
At late times only the growing mode survives and hence we must choose
the initial density and velocity field to put the system in the
growing mode.
Density contrast $\delta$ is related to $\phi$ through
eqn.\ref{poisson}, and in linear theory the velocity field can also be 
expressed in terms of the gravitational potential $\phi$ in the
growing mode. 
Therefore our problem reduces to generating the gravitational
potential and using it to set up the density contrast and velocity
field. 
The relation between the velocity and the potential in Zel'dovich
approximation \cite{za} is the same as that in the linear theory,
therefore it is often said that the Zel'dovich approximation is used
to set up the initial conditions for N-Body simulations \cite{init0}.

Given an initial gravitational potential $\phi_{in}$, there are two
schemes for generating the initial density field.

\begin{itemize}
\item  
The particles are distributed uniformly and their masses are chosen
so that $M = {\bar\rho}(t_{in}) ( 1 + \delta_{in} ) dV$, where $\delta_{in}$ 
is evaluated at the position of the particle.  Here $dV$ is the volume
in the simulation box per particle and ${\bar\rho}(t_{in})$ is the
average density at time $t_{in}$.
We can either start with zero velocities, in which case we have to
increase the amplitude of $\phi_{in}$ by a factor $5/3$ to account for
the presence of decaying mode.  
Alternatively we can choose to put the system in the growing mode and
assign the appropriate velocity to each particle.  
\item  
Starting with a uniform distribution the particles are displaced using
the velocity in linear theory for the growing mode.
It is important to ensure that the maximum displacement is smaller
than the average inter-particle separation in the simulation box. 
It can be shown that the resulting distribution of particles will
represent the required density field \cite{jbtp_pm} if the initial
distribution did not have any inhomogeneities.
We can retain the initial velocity field used to displace particles.  
If the amplitude of displacements used is larger than the average
inter-particle separation, it becomes necessary to recompute the
potential from displaced positions and assign initial velocities with
this potential \cite{p3m}. 
Such large displacements can lead to an incorrect realisation of the
power spectrum.
\end{itemize}

Schemes outlined above require an initial uniform distribution of
particles.  
This is important as any inhomogeneities present in the initial
distribution will combine with the density perturbations that are
generated by displacing particles and will modify the initial
conditions. 
\begin{itemize}
\item
The commonly used solution is to place particles on a cubic grid, this
is a uniform but not a random distribution.  
\item
An intuitive solution is to put particles at random inside the
simulation box.  
This distribution has $\sqrt n$ fluctuations which result in spurious
clustering that tend to dominate over the fluctuations we wish to
simulate. 
\item
Particles are placed in lattice cells but at a random displacement
from the centre of the cell \cite{jbtp_pm}.  
This removes the regularity of grid without sacrificing uniformity.  
The amplitude of fluctuations can be controlled by reducing the
amplitude of displacement about the centre of the cell.
\item
The {\it Glass} initial conditions are obtained by evolving an arbitrary
distribution of particles in an N-Body simulation with a repulsive force.  
It can be shown that the amplitude of perturbations oscillates and
decreases as $a^{-1/4}$ \cite{jbtp_pm}.
\end{itemize}

\subsubsection{The Initial Gravitational Potential}

The initial density field is taken to be a Gaussian random field in
most models. 
Linear evolution does not modify the statistics of density fields
except for evolving the amplitude of perturbations.  
As the potential and density contrast are related through a linear
equation, it follows that the gravitational potential is also a
Gaussian random field.
A Gaussian random field is completely described in terms of its power
spectrum \cite{bbks}.  
The Fourier components of a Gaussian random field (both the real and
the imaginary part) are random numbers with a normal distribution
with variance proportional to the power spectrum of the random field. 
This property is used to generate the Gaussian random field in
Fourier space and an inverse transform gives us the initial potential
in real space. 

If some special features are required in the initial conditions, e.g.,
if we want a large planar perturbation, then we need to impose constraints on
the Gaussian random field to be generated \cite{init1,init2,init3,init4}.  
Gaussian random fields with a variable resolution \cite{multig} are
needed for adaptive mesh refinement codes.


\section{Limitations of N-Body Simulations}

Here we discuss the scope and limitations of N-Body simulations.  
We will consider several issues concerning the domain of validity for
N-Body simulations.  

N-Body simulations take initial density fluctuations over a finite range of
scales into account.  
Do the fluctuations that are not taken into account make a difference to
results of N-Body simulations?  

Several N-Body studies have shown that fluctuations at small scales do not
affect structure that forms at large scales
\cite{little,trpwr,renorm1,renorm2} in a significant manner. 
Effects are of course there if larger scales are in linear regime but by the
time larger scales reach $\delta \sim 1$, influence of smaller scales is
ignorable. 
These studies used power spectrum, correlation function and visual appearance
to reach this conclusion.  
Therefore in any N-Body simulations that are used, the scale where
root mean square fluctuations are unity should be clearly resolved for
results to be reliable and independent of the small scale cutoff. 
On the other hand we expect some effect of small scale fluctuations on how
larger density perturbations relax \cite{previr,noise,ma,kin_halo,planar1}. 
We can conclude that small scale fluctuations do not influence large
scales in a significant manner but it is an important issue and
further studies are required to make the effect or the lack of it more
quantitative.  

\begin{figure}
\includegraphics[width=3.4truein]{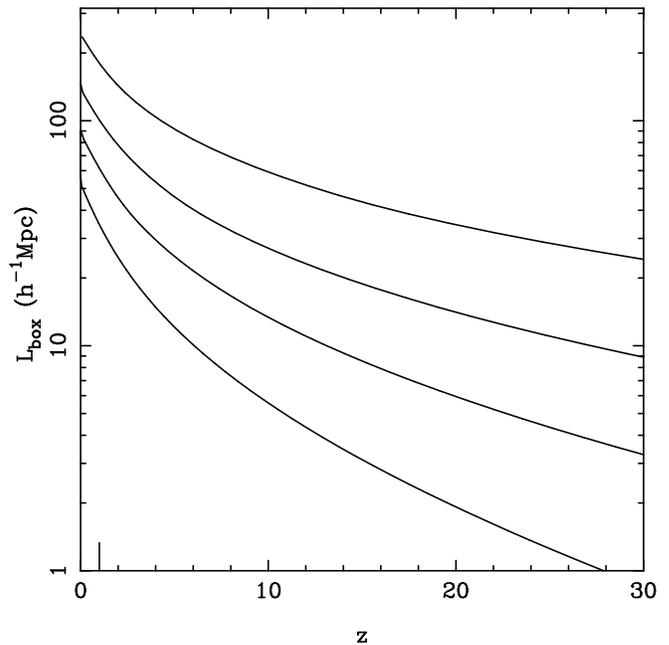}
\caption{Lines of constant root mean square fluctuations in mass
  $\sigma$ are plotted as a function of scale and redshift for the
  $\Lambda$CDM model (see text for details).  Curves are for
  $\sigma=0.1$, $0.05$, $0.025$ and $0.01$ in ascending order.  We can
  find out the lowest redshift up to which a given simulation box may
  be used once we fix $\sigma$ that is considered acceptable at the
  box size.  Such a threshold does not carry any information about the shape
  of the power spectrum.  Comparing with an approach based on convergence of
  collapsed mass, we find that $\sigma=0.025$ is a reasonable choice
  at low redshifts \cite{nbody_lims}.}
\label{box_lims}
\end{figure}

N-Body simulations assume that the density in the simulation box is same as
the average density in the universe.  
Therefore we must choose a simulation box such that the amplitude of
fluctuations in the universe (or the model being simulated) at that
scale is ignorable.   
Studies have shown that violating this requirement leads to an underestimate
of correlation function though the mass function of small mass haloes
does not change by much \cite{gelb_bert1,gelb_bert2}.
Effects at large scales can be significant \cite{void_large,berkloeb}. 
In other words, the formation of small haloes is not disturbed but their
distribution is affected by non-inclusion of long wave modes. 
One way of quantifying fluctuations at large scales, and hence their
effect on structure formation at small scales is the amplitude of
fluctuations at the scale of the simulation box. 
Figure~\ref{box_lims} shows lines of constant root mean square
fluctuations in mass $\sigma$.
These lines are plotted as a function of scale and
redshift for the $\Lambda$CDM model ($\Omega_\Lambda=0.7$,
$\Omega_B=0.05$, $h=0.7$ and $n=1.0$). 
Curves are for $\sigma=0.1$, $0.05$, $0.025$ and $0.01$ in ascending
order. 
We can find out the lowest redshift up to which a given simulation box
may be used once we fix $\sigma$ that is considered acceptable at the
box size.  

However, a threshold in $\sigma$ does not carry any information about the
shape of the power spectrum and that is extremely relevant here.
We find that the best way of quantifying the effect of long wave modes is to
check whether including them in the simulation will change the number of
massive haloes or not \cite{nbody_lims}.
This can be estimated using the Press-Schechter mass function \cite{ps75}. 
If there is a measurable effect on the number of haloes and collapsed
mass, the long wave modes are relevant and must be considered.
The large scale structure in N-Body simulations does not converge until all
such modes are taken into account \cite{nbody_lims}.  
Our results for the  $\Lambda$CDM model are:
\begin{itemize}
\item
A box size of $150$h$^{-1}$Mpc is needed for simulations that are evolved to
the present epoch ($z=0$). 
\item
Simulations that are evolved up to $z \sim 3$ should be $50$h$^{-1}$Mpc across
for generic applications.
\item
Simulations for studying early reionisation ($z \simeq 10$) should be at least
$20$h$^{-1}$Mpc across if these are to have a representative density field.
\end{itemize}
Comparing these with the curves in figure~\ref{box_lims}, we find that
$\sigma=0.025$ is a reasonable choice except at very high redshifts. 
$\sigma=0.01$ is a better choice at higher redshifts, and this variation in
the threshold value is indicative of the dependence on the shape of the power
spectrum \cite{nbody_lims}.

The effect of a finite simulation box on modes comparable with the box size
can be estimated analytically \cite{box_mod} and this can also be used to
check whether the chosen physical size of the simulation box is sufficiently
large or not. 

Methods have been developed to take the missing long wave modes into
account \cite{longw1,longw2}, these are wave modes larger than the
simulation box.   
These methods evolve a realisation of long wave modes independently
and their effect is combined with the evolution of small scales in the
N-Body simulation.  
This allows one to work with smaller simulation boxes, but the methods
have limitations and cannot be used for simulating very small regions. 
Indeed, the criterion based on convergence of collapsed mass can be
used to fix the smallest box size for which these methods can be
employed \cite{nbody_lims}.

Other limitations relate to two body relaxation, poor accuracy in force
calculation and discreteness.  
It has been asserted that in collisionless simulations forces are not
required to be very accurate as the discreteness noise dominates over
errors in force \cite{discr1,discr2}. 
This is more relevant for non-cosmological simulations, where many of
these tests were carried out. 
More recent studies of force softening have pointed out that the form of
softening can be very important \cite{optsoft1} in controlling the
discreteness noise.  

Tests that are specific for cosmological simulations \cite{discr3}
show that unless mass and force resolution are comparable, two body
relaxation can leave detectable signatures. 
Thus it is important to avoid discreteness effects and maintain parity
between mass and force resolution in N-Body codes. 


\section{Non-linear Gravitational Clustering}

N-Body simulations have been used extensively to understand aspects of
gravitational clustering.  
Of the four fundamental interactions, gravity is the weakest and
it is impossible to carry out laboratory experiments in
gravitational dynamics of a many body system as other interactions
overwhelm gravity at these scales.  
N-Body simulations can be used as a test-bed for doing numerical
experiments in order to understand various aspects of non-linear
gravitational clustering. 
Indeed, we cannot claim to have understood the process of structure
formation till we develop sufficient insight so as to make N-Body
simulations redundant except for the purpose of computing detailed
predictions for models of structures formation. 

Considerable progress has been made using quasi-linear approximations,
non-linear scaling relations and N-Body simulations, though many
important questions are yet to be settled. 
Basic premise here is that if we are interested in only a limited
amount of information about the final state then it should be possible
to simplify the calculation by making an ansatz that captures the
essential physical process at work. 
One can also invert this and find out which physical process dominates
in a given situation. 
We review some of the issues that have been studied in detail. 

\medskip

\noindent
The distribution of matter is very close to homogeneous at early
times, how is it confined to central region of potential wells at late
times?  
\begin{itemize}
\item
Infall is described very well by first order Lagrangian perturbation
theory. 
This, extrapolated to mildly non-linear density contrasts is called
the Zel'dovich approximation \cite{za}.
This approximation suggests that the first structures to form as a
results of collapse are typically surfaces of high density, the so
called pancakes.
Comparisons with N-Body simulations show this to be a very good
approximation up to this stage \cite{tza}. 
However, these surfaces of high density thicken and disappear if we
extrapolate the approximation scheme to late times, whereas these
pancakes do not thicken in N-Body simulations. 
Clearly, the Zel'dovich approximation and even higher order Lagrangian
perturbation theory \cite{lagpert1,lagpert2,lagpert3} lacks the key
ingredient that helps confine matter to potential wells.
\item
An artificial viscosity term can be added to the equation of motion,
eqn.(\ref{motion}), and it can be transformed into Burger's equation
\cite{adh}. 
This prevents interpenetration of infalling streams and pancakes
remain thin, this is called the Adhesion approximation
\cite{adh,adh2}. 
The approximation puts matter in the right place but fails to find a
physical reason for the viscosity term.
\item
The equation of motion, eqn.(\ref{motion}), contains a velocity
dependent term where the expansion of the universe acts as a viscous
drag. 
Further, it can be shown that the gravitational potential $\phi$
varies very slowly.  
If we assume that the potential changes at the rate expected in the
linear theory while retaining its initial spatial dependence then the
drag force is sufficient to confine matter to the central regions of
potential wells \cite{lep,fpa}. 
This approximation ({\sl Linear Evolution of Potential}\/ (LEP), {\it
  aka} {\sl Frozen Potential Approximation} \/ (FPA)) compares well
with N-Body simulations. 
We can conclude that the drag term in the equation of motion plays an
important role in confining matter to potential wells and the
interaction of infalling particles plays a less significant role.
\end{itemize}

\medskip

\noindent
How strongly do density fluctuations at small scales 
influence density perturbations at large scales?  This is relevant as
N-Body simulations ignore perturbations at scales smaller than those
probed in the simulation.
\begin{itemize}
\item
If there are no perturbations at large scales then small scale
fluctuations generate a $P(k) \propto k^4$ spectrum at small $k$, this
effect can be derived in the second order perturbation theory as well
as from the full set of equations \cite{lssu}. 
Amplitude of this limiting spectrum increases rapidly at early times
\cite{trpwr}. 
This effect of course cannot be seen if there are fluctuations at
large scales with $P(k) \propto k^n$ with $n < 4$ at small $k$. 
\item
If there are small density perturbations present at large scales
then there is a discernable visual effect of small scales on
perturbations at large scales
\cite{renorm1,renorm2,little,previr_sim,trpwr} for generic initial
conditions. 
Small scales can influence the power spectrum and higher moments of
particle distribution at these scales, the detailed effect depends on
the form of fluctuations at small scales \cite{modec2}. 
\item
If there are non-linear density perturbations present at large scales
then there is no discernable visual effect of small scales on
perturbations at large scales
\cite{renorm1,renorm2,little,previr_sim,trpwr} for generic initial
conditions. 
There is no significant effect in the power spectrum or the two point
correlation function \cite{little,trpwr} in such a case. 
If the small scales and the scale of non-linearity are well separated
then there is no effect in higher moments either \cite{modec2}.
\item
Clearly, non-linear gravitational clustering works to remove the
effect of perturbations at large scales. 
\item
For special initial conditions, it has been found that collisions
between clumps formed due to small scale perturbations can lead to a
faster relaxation of perturbations at large scales.
E.g., if larger scales are modelled as a single plane wave then the
resulting pancake is thinner in presence of significant small scale
fluctuations \cite{planar1}. 
It remains to be seen whether this is an important effect for generic
initial conditions.
\end{itemize}

\medskip

\noindent
Does gravitational clustering erase memory of initial conditions?  
Is there a one to one correspondence between some characterisation of
initial perturbations and the final state? 
Note that this is different from removing effects of perturbations at
scales much smaller than the scale of non-linearity.  
Clearly, if the answer to the first question is {\sl yes} then we
cannot recover any information about the initial density perturbations
from determination of density perturbations in the non-linear regime.
\begin{itemize}
\item
N-Body simulations show that gravitational clustering does not erase
memory of initial conditions
\cite{hamil,jmw,pd94,pd96,patterns,sm04}. 
The final power spectrum is a function of the initial power spectrum
and this relationship can be written as a one step mapping. 
\item
The functional form of this mapping depends on the initial power
spectrum \cite{sm04}.
\item
An analytical understanding of some aspects of this mapping can be
developed using simple models and approximations \cite{rntp,tp96}. 
\item
It is found that density profiles of massive haloes have a form
independent of initial conditions \cite{nfw,powerlaw,kandu2}. 
It is important to note that there is considerable scatter in density
profiles obtained from N-Body simulations and it is difficult to state
whether a given functional form is always the best fit or not.
There are a large number of recent studies (e.g. see
\cite{denp1,denp2}) but most of these focus on the $\Lambda$CDM model
and do not explore other initial conditions. 
\end{itemize}

\medskip

\noindent
Is it possible to predict the masses and distribution of haloes that
form as a result of gravitational clustering?
\begin{itemize}
\item
The initial density field is taken to be a Gaussian random field, and
for hierarchical models \cite{lssu} the simple assumption that each
peak undergoes collapse independent of the surrounding density
distribution can be used to estimate the mass function
\cite{ps75,ps_bond} and several related quantities.
\item
N-Body simulations show that this simple set of approximations is
incorrect, however the resulting mass function is fairly accurate over
a wide range of masses.
\item
Merger rates can be computed using the extended Press-Schechter
formalism \cite{ps_bond}, these are useful for many applications
\cite{ps_merg1,ps_merg2,ps_merg3}.  
It is also possible to estimate clustering properties of haloes using
this formalism \cite{mwbias,mjw}.
\item
Modifying some of these assumptions can lead to improved predictions
\cite{pgmf,massf1,massf2}. 
\end{itemize}

We have highlighted some of the key issues in non-linear gravitational
clustering in this section and reviewed how N-Body simulations and
various approximations have been used to develop insight.  
N-Body simulations have been and are being used for a variety of other
applications, the most notable being computing detailed predictions
for models of structure formation.


\section{Gastrophysical Effects}

N-Body simulations that take only gravitational interactions into
account can be used to obtain the large scale distribution of galaxies
and clusters. 
Further details can only be obtained using simulations with gas
physics. 
There are two very different approaches to including gas physics in
N-Body simulations.

The classical approach is to use a grid to solve fluid equations
\cite{evrard,zeus2d1,ppm,zeus3d,mesh_hydro,mesh_hydro0,mesh_hydro1,mesh_hydro2,ahydro,hydra0,hydra}.   
Fluid interactions are short range and information from nearby grid
points is sufficient to evolve fluid properties at any grid point.  
Of course, fluids must respond to gravitational field of the matter
distribution. 
This type of a code is well adapted for capturing shocks and
discontinuities. 
Resolution of the code can be enhanced in high density regions by
using mesh refinement.  
In cosmological simulations it is important to improve the mass
resolution of dark matter particles along with an enhancement in
resolution for hydrodynamics. 
Grid codes can be easily generalised to include other effects such as
magnetohydrodynamics \cite{zeus2d2,zeus2d3}. 
Such codes have been used effectively for cosmological applications
\cite{evrard,mesh_hydro,mesh_hydro1,ahydro,mesh_hydro2,hydra0,hydra}. 

The {\sl Smooth Particle Hydrodynamics} (SPH) \cite{sph,sphrev1} where
particles are assigned fluid properties is a very different approach
to the same problem. 
The fluid properties like pressure, density, temperature, etc. at any
point can be found by averaging over particles in the region using a
weight function. 
Use of interpolation for determining fluid properties makes it
impossible to resolve shocks and discontinuities in SPH codes.
There are several known limitations
\cite{sph_lims0,sph_lims1,sph_lims2} that one should be aware of. 
SPH codes are relatively easy to implement and several implementations
are in use.  
Several implementations for cosmological simulations are in use
\cite{grapesph,treesph,partreesph}. 
Variations of SPH with focus on entropy equation
\cite{entropy0,entropy1} or on simulation of multi-phase media
\cite{mulph1,mulph2} have been developed.

Both types of codes have been compared and give similar results
\cite{kang,cluscomp} for astrophysical applications. 

Effects other than hydrodynamics can also be incorporated in a
similar fashion in both types of codes.  
For example elementary chemical reactions like formation of Hydrogen
molecules, cooling, heating, etc. are important local effects.
Key effects like star formation and feedback from stellar and other
sources are difficult to include as vastly different scales are
of relevance.  
As a result, much of the treatment of these effects has remained
phenomenological. 

Radiation transport is a non-local effect and is difficult to take
into account
\cite{radtrans1,radtrans2,radtrans3,radtrans4,radtrans5,radtrans6,radtrans7}. 
The radiation field is either assumed to be isotropic and homogeneous,
or it is assumed to originate at a few point sources. 
The time scales over which the radiation field evolves are much
shorter than most other time scales in the problem.
Thus we can assume that the density and velocity field do not change
while studying  evolution of the radiation field.

State of the art simulations can be used for studying a variety of
problems.  We summarise three systems here:
\begin{itemize}
\item
The inter-galactic medium \cite{igm,igm0,igm1,igm2,igm3,igm4} is
believed to contain mostly pristine material with elements other than
Hydrogen and Helium present in only trace amounts.  
The radiation field can be assumed to be isotropic and homogeneous at
$z \leq 3$, though at higher redshifts it is patchy.  
Densities in the inter-galactic medium do not become very large.
\item
The intra-cluster medium
\cite{cluscomp,icm0,icm1,icm2,icm3,icm4,icm5,icm6,icm7,icm8,icm9,gadget2} is 
believed to be close to hydro-static equilibrium. 
Gas in the intra-cluster medium is so hot that many radiative
processes can be ignored safely.
There are several sources of energy in clusters of galaxies, most of
these are distributed uniformly with some contribution from a central
AGN. 
Magnetic fields may be important enough to make the physics of the
problem more complicated.  
\item
Formation of first stars
\cite{fstars1,fstars2,fstars3,fstars4,fstars5,fstars6,fstars7,fstars8}
requires simulating cooling of Hydrogen (mainly) through various
processes till molecules form and lead to temperatures low enough for
formation of stars. 
\end{itemize}

Focus in future will be on incorporating further effects so that more
complex problems like galaxy formation can be studied in full detail.
N-Body simulations are useful not only as tools for evolving complex
systems, these can also be used to understand which effects play a
more important role in different phases of this evolution. 
Next decade holds the promise of understanding physics of galaxy
formation with help of N-Body simulations.


\section*{Acknowledgements}

Developing insight into developments in cosmological N-Body simulations would
not have been possible without access to a high performance computing
facility.  
Cluster computing facilities at the Harish-Chandra Research Institute
(http://cluster.mri.ernet.in) were used heavily for trying, testing and also
developing some of the algorithms.
I thank Suryadeep Ray, with whom I have tried out many experiments
with N-Body algorithms.
I also thank H.K.Jassal and Suryadeep Ray for a careful reading of the
manuscript. 
This research has made use of NASA's Astrophysics Data System.


\end{document}